\documentclass[prx,final,twocolumn,superscriptaddress,aps]{revtex4-2}
\usepackage{amsmath}
\usepackage{amssymb}
\usepackage{bbold}
\usepackage[utf8]{inputenc}
\usepackage[colorlinks=True,linkcolor=blue,citecolor=blue,urlcolor=blue]{hyperref}
\usepackage{xcolor}
\usepackage{graphicx}
\usepackage{physics}
\usepackage{times}

\begin{document}
\title{Scattering States in One-Dimensional Non-Hermitian Baths}
\author{Jimin Li}
\affiliation{Department of Applied Mathematics and Theoretical Physics, University of Cambridge, Wilberforce Road, Cambridge CB3 0WA, United Kingdom}
\author{Yuwen E. Zhang}
\affiliation{Department of Physics and Astronomy, University College London, Gower Street, London WC1E 6BT, United Kingdom}
\author{Franco Nori}
\affiliation{Theoretical Quantum Physics Laboratory, Cluster for Pioneering Research, RIKEN, Wako-shi, Saitama 351-0198, Japan}
\affiliation{Quantum Computing Center, RIKEN, Wakoshi, Saitama 351-0198, Japan}
\affiliation{Department of Physics, University of Michigan, Ann Arbor, Michigan 48109-1040, USA}
\author{Zongping Gong}
\affiliation{Department of Applied Physics, University of Tokyo, 7-3-1 Hongo, Bunkyo-ku, Tokyo 113-8656, Japan}

\begin{abstract}
{A single quantum emitter coupled to a structured non-Hermitian environment shows anomalous bound states and real-time dynamics without Hermitian counterparts, as shown in [Gong \emph{et al.}, Phys. Rev. Lett. 129, 223601 (2022)]. In this work, we establish a general approach for studying the scattering states of a single quantum emitter coupled to one-dimensional non-Hermitian single-band baths. We formally solve the exact eigenvalue equation for all the scattering states defined on finite periodic lattices. In the thermodynamic limit, the formal solution reduces to the celebrated Lippmann-Schwinger equation for generic baths. In this case, we find that the scattering states are no longer linear superpositions of plane waves in general, unlike those in Hermitian systems; Instead, the wave functions exhibit a large, yet finite localization length proportional to the lattice size. Furthermore, we show and discuss the cases where the Lippmann-Schwinger equation breaks down. We find the analytical solutions for the Hatano-Nelson and unidirectional next-to-nearest-neighbor baths in the thermodynamic limit. 
} 
\end{abstract}
\maketitle
\section{Introduction}
\label{sec:Introduction}
\begin{figure}[t]
  \centering
  \includegraphics[width=1\columnwidth]{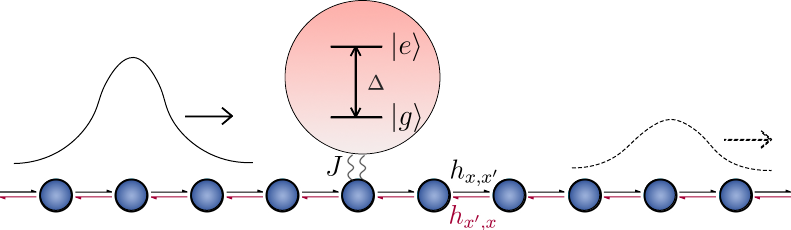}
  \caption{ Illustration for a single quantum emitter (two-level) coupled to 1D nearest-neighbor hopping NH baths with $|x' - x| = 1$. Other structured baths are also studied in this work. The wave indicates that the propagation of the photon is scattered by the emitter.
  }  
  \label{fig:fig1}
\end{figure}

Understanding light-matter interactions has been the central problem in modern quantum platforms in order to realize controllable quantum optical systems for fundamental studies and practical applications \cite{Chang2018,Blais_review,chang_quantum_2014,Calaj02016,kannan_waveguide_2020,Zheng13,Zhang2021,KockumDecoherence2018, DongWaveguide2025,Johansson2009dynamical,Selim_2025,selim2025engineeringnonhermitianquantumevolution}. A prototypical setup is to consider quantum emitters, which are atoms consisting of a few discrete energy levels coupled to structured baths. Such structured baths are an effective description of a large number of degrees of freedom and model a range of open quantum systems. A paradigmatic example is the study of emitters coupled to baths with waveguiding structures, also known as waveguide QED \cite{RevModPhys.95.015002, Dong2025waveguide, Terradas2022ultra, huang2013controlling, li2018quantum, nie2023non-hermitian}. In this context, there have been many studies concerning phenomena like bound states, transmission rate, and real-time photon dynamics, which are highly relevant in realizing practical quantum optical usages such as photon devices and quantum information storage. Many of the properties are closely related to the dispersion relations of the structured baths, and engineering more sophisticated structured baths, such as topological photonic baths, can lead to novel effects \cite{Bello2019,Leonforte2021,Vega2023}. 

In recent years, variants of structured baths have been enriched by the advent of non-Hermitian (NH) physics \cite{Wang:24, UedaReview, Poddubny2024}. The NH systems alone show many properties without Hermitian counterparts; for example, the dispersion is complex in general and supports non-trivial winding numbers for one-dimensional systems \cite{Gong2018, KawabataPRX2019}. Such peculiar features have been reported for quantum emitters coupled to NH baths, such as novel bound states \cite{GongPRL2022, cai2024chiral}, topological reversal \cite{roccati2024hermitian}, fractional Zeno effect \cite{Sun2023}, and anomalous collective spontaneous emission \cite{Li2024}, to name a few. Many of the exciting NH physics have been realized experimentally \cite{leefmans2022topological, parto2023non,leefmans2024topological,jin2023reentrant}.

Yet the manifestation of NH physics in single-photon scattering off a quantum emitter has been overlooked. Scattering processes are ubiquitous in optical platforms, where the interaction between the incident photons and the scatterer produces non-trivial coherent transport of photons. For single-photon dynamics in Hermitian baths, these scattering processes have been explored on variants of circuit QED platforms \cite{Astafiev2010, lu2014single} and applied to quantum devices such as quantum switches \cite{Zhou2008, zhou2009quantum}, atomic mirrors \cite{zhou2008quantum}, and cavities \cite{liao2010controlling}. This work aims to generalize the single-photon scattering to NH baths. From a theoretical perspective, the simplest model considers the propagation of a photon in the structured baths scattered off by a quantum emitter (see Fig.~\ref{fig:fig1}). The real-time dynamics of the photon can be tracked by considering the spectral decomposition of the underlying Hamiltonian, which involves knowledge of the exact eigenvalues and eigenvectors. The main focus of this work is to investigate the extensive eigenvectors for a single quantum emitter coupled to one-dimensional NH baths. These eigenvectors exhibit (almost) the same eigenvalues as the baths and are usually referred to as scattering states.

Note that while the same NH setup is studied in Refs.~\cite{GongPRL2022, GongPRA2022}, the focus therein is on bound states and real-time dynamics; the scattering wavefunction remains largely unexplored. Our setup also differs fundamentally from the previous studies on scattering off a local non-Hermitian potential \cite{Ali2009}, since here non-Hermiticity prevails over the whole space.

The theoretical description of scattering states has been established since the early days of quantum mechanics. It is well-known that the Lippmann-Schwinger (LS) formalism \cite{Lippmann1950} provides a systematic approach to describing general scattering processes ranging from standard quantum mechanics to field theory, including those on quantum optical platforms \cite{Shen2005,Shen2007}. However, the validity of the LS formalism for describing scattering states in NH baths remains unclear.

Indeed, in a direct application of the standard LS equation to NH systems, one immediately sees technical difficulties in choosing the correct analytically continued branch of the Green's function. In Hermitian systems, this issue is solved by adding a tiny imaginary shift to the energy. However, this solution relies heavily on the fact that the spectra of Hermitian Hamiltonians are real. It is not clear whether, and if yes, how a generalization to non-Hermitian systems could be achieved.

In this work, we address the above question. We derive the formal expression for the scattering wavefunction on finite lattices from the exact eigenvalue equation. And we show that the formal solution reduces to the LS formalism in the thermodynamic limit for generic baths. Furthermore, the LS formalism can also break down in fine-tuned situations. Two minimal models of 1D NH baths are considered as concrete examples, which are NH nearest-neighbor (NN) hopping \cite{Hatano1996} and unidirectional NH next-to-nearest-neighbor (NNN) hopping models. We find the analytical expressions for all the scattering states in the thermodynamic limit. And all the bound states (including the anomalous ones) are obtained by complexifying the real momentum in the appropriate scattering states. In general, the scattering states are not a linear superposition of plane waves, in contrast to the standard scattering problem in (Hermitian) quantum mechanics. In addition, we also find plane wave scattering states in NH baths with fine-tuned spectral properties, where the LS approach breaks.

The rest of the paper is structured as follows. We first give the definition of the model in Sec.~\ref{sec:Setup} and subsequently present our general approach to the scattering states in Sec.~\ref{sec:Formal-Solution}. In Sec.~\ref{sec:HN} (\ref{sec:NNN}), we present the analytical results for the LS wavefunctions for the HN (NNN) baths, and fine-tuned cases are discussed in Sec.~\ref{sec:fine-tuned-points}. Finally, our results are summarized in Sec.~\ref{sec:conclusion}.

\section{Model}
\label{sec:Setup}

We consider a single
quantum emitter at site $x=0$, modeled by a
two-level atom with ground (excited) state $|g\rangle$ ($|e\rangle$), coupled to a 1D dissipative bath consisting of $L$ sites via a Jaynes-Cummings interaction \cite{CohenAP}. The dissipative baths are engineered to have the desired NH dispersion by changing the hopping parameters and range, see Ref. \cite{GongPRL2022} for an explicit realization of the NH baths by post-selection of the corresponding Lindbladian. Using the full Lindbladian treatment in the approximation regimes where it is valid (i.e., Markovian and rotating-wave) \cite{GongPRA2022}, the effective NH Hamiltonian reads 
\begin{equation}
    H = H_0 + V,
\label{eq:Ham-realspace-full}
\end{equation}
where 
\begin{equation}
    H_0 = \Delta |e \rangle \langle e | + \sum_{x,x'} h_{x,x'} a^{\dagger}_{x}a_{x'}
\label{eq:Ham-realspace}
\end{equation}
with $a_x$ ($a^\dag_x$) being the photon annihilation (creation) operator on site $x$. Here the first term in Eq.~(\ref{eq:Ham-realspace}) is the Hamiltonian for the quantum emitter, and $\Delta$ is the detuning, i.e., energy gap in the rotating frame.
The second term is the bath Hamiltonian and the details of the bath are fixed by parameters $h_{x,x'}$.  
The coupling term 
\begin{equation}
    V = J ( |g \rangle \langle e |a^{\dagger}_{x=0} + a_{x=0} |e \rangle \langle g| )
\end{equation}
is known as the Jaynes-Cummings coupling and $J$ is the coupling strength. We emphasize again that here the baths are intrinsically NH, and the coupling is Hermitian. This setup is essentially different from Hermitian background scattering off NH potentials, e.g., studied in Ref. \cite{Ali2009}.

In this work, we focus on translation-invariant baths. In particular, periodic boundary condition (PBC) is imposed for the 1D baths of length $L\in\mathbb{Z}^+$, i.e., a ring of length $L$. This can be implemented by, e.g., superconducting circuits with flexibly engineerable architecture \cite{PhysRevApplied.16.024018}. We assume the NH baths have a left and right hopping range $p$ and $q$ with the corresponding hopping strength given by $h_{x,x'}$. Throughout the paper, we consider a sufficiently large bath with $L\gg p+q$. Under PBC, we use $n= x-x'$ to label the relative hopping distance (i.e., $h_{x,x'}=h_{x-x'}=h_n$) and the NH band dispersion reads
\begin{equation}
    h_k=\sum_{n=-p}^q h_{n} e^{-ink},
    \label{eq:hk}
\end{equation}
where $k=2\pi m/L$ with $m=1,2,...,L$. 

Overall, the effective NH Hamiltonian Eq.~(\ref{eq:Ham-realspace-full}) has the following form in the momentum space
\begin{equation}
    H%(k) 
    = \Delta |e\rangle\langle e| + \sum_{k} h_k a^{\dagger}_{k} a_k + \frac{J}{\sqrt{L}}\sum_k \left( |g\rangle \langle e | a^{\dagger}_k + a_k |e \rangle \langle g|   \right),
\label{eq:NH-Ham-k}
\end{equation}
where 
\begin{equation}
    a_k = \frac{1}{\sqrt{L}} \sum_{x} e^{-ik x } a_{x}.
\end{equation}

While we focus primarily on PBC, let us briefly comment on the open boundary condition (OBC). For NH systems, boundary conditions are widely recognized to have a non-trivial effect on the physics \cite{Yao2018,li2024dissipation,UedaReview}. For our setup under the OBC, all the scattering states and anomalous bound states are expected to be localized at one end, while the conventional bound states remain localized around the emitter. This observation has been mentioned in Ref.~\cite{GongPRA2022}, and has been numerically confirmed for all the models in this work.

\section{Formal Solution}
\label{sec:Formal-Solution}
\begin{figure}[t]
  \centering
  \includegraphics[width=0.9\columnwidth]{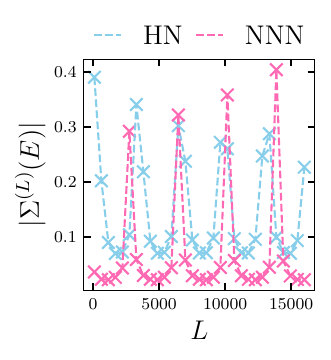}
  \caption{The magnitudes of finite-size self-energy Eq. (\ref{SL}) evaluated for the HN and unidirectional NNN baths. For the same randomly selected $E = h_{k}$, the finite-size self-energies do not converge with the system size.} 
  \label{fig:fig0}
\end{figure}

The simplest scattering setup is to consider the single excitation sector of the NH Hamiltonian defined by Eq.~(\ref{eq:NH-Ham-k}), which has the following exact eigenvalues and eigenvectors
\begin{equation}
    H |\Psi _i \rangle = E_i |\Psi_i\rangle. 
\label{eq:exact-eigen}  
\end{equation} 
We consider a non-vanishing coupling strength $J$ since otherwise the solutions are simply Bloch waves and the emitter excitation. 

Any state in the single excitation sector is given by {an expansion under the complete basis:
\begin{equation}
    |\phi \rangle = \left( c_e |e\rangle \langle g| + \sum_k \frac{c_k}{\sqrt{L}}a^{\dagger}_k \right) |g \rangle \otimes | \text{vac} \rangle, 
\label{eq:complete-basis}
\end{equation}
where $c_{k,e}$ are coefficients to be determined and $| \text{vac} \rangle$ denotes the photon vacuum.
While usually not mandatory in scattering problems, the normalization condition reads
\begin{equation}
    |c_e|^2+L^{-1}\sum_k|c_k|^2=1.
\label{eq:Normalization}
\end{equation}
Substituting this complete basis expansion (\ref{eq:complete-basis}) into Eq.~(\ref{eq:exact-eigen}) leads to the following more explicit eigenvalue problem:
\begin{align}
    \begin{split}
        \Delta c_e + \frac{J}{L} \sum_k c_k &= E_i c_e, \\
    h_k c_k + J c_e &= E_i c_k, \forall k=2\pi m/L.
    \end{split}
\label{eq:Eigenvalue-conditions}
\end{align}

Here the central object is the photon state, $|\psi\rangle =\sum_k c_k|k\rangle / \sqrt{L}$ as it represents the scattering (bound) state in the discrete waveguide model if its real-space profile is extensive (localized). Our goal here is to find a formal solution for the photon wavefunction in real space
\begin{equation}
\psi(x)=\langle x|\psi\rangle=\frac{1}{\sqrt{L}}\sum_k c_k e^{ikx},
\end{equation}
where $x=-\lfloor L/2\rfloor ,...,-1,0,1,...,\lfloor (L-1)/2\rfloor$ is defined on $\mathbb{Z}_L$, i.e., $x+L$ and $x$ should be identified under PBC.

Under the assumption of $c_e \neq 0$, the exact eigenvalue equation Eq.~(\ref{eq:Eigenvalue-conditions}) is equivalent to 
\begin{align}
    \begin{split}
        E-\Delta-\Sigma^{(L)}(E)&=0, \\
    c_k=\frac{J}{E-h_k} c_e,
    \end{split}
\label{eq:Eigenvalue-conditions_ceneq0}
\end{align}
where $\Sigma^{(L)}(z)$ is the finite-size ($L$-site) self-energy given by
\begin{equation}
\Sigma^{(L)}(z)=\frac{J^2}{L}\sum_k\frac{1}{z-h_k},
\label{SL}
\end{equation}
and $z \in \mathbb{C}\backslash\{h_k\}_k$. By introducing its generalization, 
\begin{equation}
\Sigma^{(L)}_x(z)=\frac{J^2}{L}\sum_k\frac{e^{ikx}}{z-h_k},
\label{SLx}
\end{equation}
which reproduces the self-energy for $x=0$, we can rewrite the photon eigen-wavefunction as
\begin{equation}
\psi(x)=\frac{1}{\sqrt{L}}\sum_k \frac{Je^{ikx}}{E-h_k}c_e = \frac{\sqrt{L} c_e}{J}\Sigma^{(L)}_x(E),
\label{pxL}
\end{equation}
provided that $E$ is a solution to $E-\Delta-\Sigma^{(L)}(E)=0$. We are particularly interested in the solutions near $h_k$, which correspond to the scattering states.

Before proceeding, we first point out some useful mathematical properties of the finite-size self-energy Eqs.~(\ref{SL}) and (\ref{SLx}) and their thermodynamic-limit counterparts:
\begin{align}
    \begin{split}
     \Sigma(z)&=J^2\int^{\pi}_{-\pi}\frac{dk}{2\pi}\frac{1}{z-h_k}, \\
    \Sigma_x(z)&=J^2\int^{\pi}_{-\pi}\frac{dk}{2\pi}\frac{e^{ikx}}{z-h_k},
    \end{split}
\label{TL}
\end{align}
where $L\to\infty$ is taken so that the sums in Eqs.~(\ref{SL}) and (\ref{SLx}) are replaced by integrals. Given a finite $L$, $\Sigma^{(L)}(z)$ and $\Sigma^{(L)}_x(z)$ are meromorphic functions of $z$, that are well-defined almost everywhere except for a set of discrete energies (NH band dispersion) $\{h_k\}_k$.

In contrast, $\Sigma(z)$ and $\Sigma_x(z)$ in Eq.~(\ref{TL}) are analytic within each connected region bounded to the (continuous) band dispersion $h_k$ ($k\in(0,2\pi]$), but generally undergo sudden jumps across the boundaries. Moreover, unlike their finite-size counterparts (Eqs.~(\ref{SL}) and (\ref{SLx})), they are \emph{not} well-defined right on the band dispersion. This fact can be examined by checking the $L$-dependence of $\Sigma^{(L)}(E)$ for some $E=h_k$ ($k\in(0,2\pi]$): in general we would observe a strong oscillation that never converges (see Fig. \ref{fig:fig0}). On the other hand, we do have a convergence for any $E$ apart from $h_k$ ($k\in(0,2\pi]$). It is thus safe to use Eq.~(\ref{TL}) to deal with a bound state whose eigen energy $E$ is well-separated from the band dispersion. 

Then how can we express a scattering state in terms of Eq.~(\ref{TL})? A quick solution is to add a small perturbation such that $E$ slightly deviates from the band dispersion, and thus the self-energies (\ref{TL}) in the thermodynamic limit become well-defined. Equivalently, we can specify a branch of analytically continued self-energy followed by substituting an energy exactly located on the band dispersion. In the following, we will show that typically it does not really matter how we perturb $E$, or which branch we choose. At first glance, the choice of branch leads to different self-energies and scattering wavefunctions. But we will show that this ambiguity can be resolved as long as the calculations are done consistently. We will also discuss atypical cases to which the consistency relation no longer applies.

\subsection{Residue Formula}
The key point is the residue formula for the finite-size self-energy $\Sigma^{(L)}(E)$ and its generalization $\Sigma^{(L)}_x(E)$, making it clear how the finite-size result is related to the thermodynamic-limit counterparts $\Sigma(E)$ and $\Sigma_x(E)$. Our approach is inspired by the Appendix of Ref.~\cite{Hatano1997}. We first focus on $x\ge0$ (recall that $x\le\lfloor L/2\rfloor$). Introducing $h(\beta)=\sum_{n=-p}^q h_n\beta^{-n}$ (such that $h_k=h(e^{ik})$), 
we have
\begin{equation}
\oint_{|\beta|=R} \frac{d\beta}{2\pi i \beta} \frac{\beta^x}{(E- h(\beta))(\beta^L-1)}=0
\label{xL}
\end{equation}
for sufficiently large $R$ such that the poles of the integrand are all located inside $|\beta|=R$. This can be seen from taking the limit $R\to\infty$ and the assumption that $L$ is large (i.e., $L\gg p+q$). According to the residue theorem, Eq.~(\ref{xL}) implies
\begin{equation}
\frac{1}{L}\sum_k \frac{e^{ikx}}{E-h_k} + \sum_{y : E=h(y)}\frac{y^x}{-y h'(y) (y^L - 1)}=0,
\label{eq:simple-poles}
\end{equation}
provided that $E$ is excluded from the finite-size spectrum $\{h_k\}_k$, $E=h(y)$ has no degenerate roots, and that $0$ is not a pole of the integrand (especially not a pole of $1/[\beta(E-h(\beta))]$). Likewise, if $x<0$ (recall that $x\ge-\lfloor (L-1)/2\rfloor$), we start from
\begin{equation}
\oint_{|\beta|=R} \frac{d\beta}{2\pi i \beta} \frac{\beta^{-x}}{(E- h(\beta^{-1}))(\beta^L-1)}=0,
\label{mxL}
\end{equation}
leading to 
\begin{equation}
\frac{1}{L}\sum_k \frac{e^{ikx}}{E-h_k} + \sum_{y : E=h(y^{-1})}\frac{y^{-x}}{y^{-1} h'(y^{-1}) (y^L - 1)}=0.
\label{eq:simplepoles2}
\end{equation}
Here, we have again assumed that $E$ is not a discrete eigenvalue, $E=h(y)$ has no degenerate roots, and $0$ is not a pole of $1/[\beta(E-h(\beta^{-1}))]$ (degeneracy of roots is actually not a problem as the higher-order residue will naturally arise from the L'H\^{o}pital rule, and the case of $0$ being a pole is considered in the end of this subsection). All together, we end up with
\begin{align}
    \begin{split}
        \Sigma_x^{(L)}(E)&=J^2\sum_{y: E=h(y)}f^{(L)}_{s_x}(y) \frac{y^{x-1}}{h'(y)}, \\
    f^{(L)}_\pm(y)&=\pm \frac{1}{y^{\pm L} - 1},
    \end{split}
    \label{SR}
\end{align}
where $s_x=+$ if $x\ge0$ and otherwise $s_x=-$ if $x<0$. As shown in Appendix~\ref{RF}, actually we can get the same result at $x=0$ (i.e., the self-energy) using the minus case involving $f^{(L)}_-(y)$:
\begin{align}
    \begin{split}
        \Sigma^{(L)}(E) &=J^2\sum_{y: E=h(y)}f^{(L)}_+(y) \frac{1}{yh'(y)} \\
    &= J^2 \sum_{y: E=h(y)}f^{(L)}_-(y) \frac{1}{yh'(y)}.
    \end{split}
\label{eq:self-energy1}
\end{align}
Also, the right-hand side in Eq.~(\ref{SR}) is indeed periodic in $x$ with period $L$.

Let us turn to compare Eq.~(\ref{SR}) with the more familiar thermodynamic-limit results. As emphasized previously, we assume $E$ is away from the continuous dispersion relation $h_k$ ($k\in(0,2\pi]$) for well-definedness. Following the standard contour integral calculations, we obtain
\begin{align}
\begin{split}
\Sigma_x(E) =\begin{cases}
        -J^2\sum_{y: E=h(y), |y|<1}\frac{y^{x-1}}{h'(y)} & x\ge0 \\
        J^2\sum_{y: E=h(y), |y|>1}\frac{y^{x-1}}{h'(y)} & x<0.
    \end{cases}
\end{split}
\label{Sx}
\end{align}
In particular, the self-energy in the thermodynamic limit reads
\begin{align}
    \begin{split}
       \Sigma(E) &= -J^2\sum_{y: E=h(y), |y|<1}\frac{1}{yh'(y)} \\
    &= J^2\sum_{y: E=h(y), |y|>1}\frac{1}{yh'(y)}.
    \end{split}
\label{SE}
\end{align}

This is fully consistent with Eq.~(\ref{SR}), since for any $|y|\neq 1$, we have
\begin{equation}
\lim_{L\to\infty} f^{(L)}_\pm(y) = \mp \theta(\mp|y|\pm1),
\label{finf}
\end{equation}
where $\theta(\;\cdot\;)$ is the Heaviside step function. Moreover, suppose $E$ is away from the continuous band dispersion, $\Sigma_x(E)$ should approximate $\Sigma^{(L)}_x(E)$ exponentially well (in terms of $L$), since $f^{(L)}_\pm$ approximate the Heaviside step functions Eq.~(\ref{finf}) exponentially well.

Lastly, we come back to the special case of $0$ being a pole of either $1/(\beta(E-h(\beta)))$ or $1/(\beta(E-h(\beta^{-1})))$. This situation only appears in the unidirectional hopping baths, where the hopping range $p=0$ or $q=0$. For $x \neq 0$, Eqs.~(\ref{eq:simple-poles}) and (\ref{eq:simplepoles2}) remain valid as the pole can be absorbed into the numerator. 
Furthermore, we make the following consistent choice: the self-energy at $x=0$ takes either the positive or the negative half expression depending on the unidirectionality of the baths. Assuming $0$ is not a higher-order pole, the self-energy reads as the positive (or negative) half expression for $p=0$ (or $q=0$), i.e., the first (or second) line of Eqs. (\ref{eq:self-energy1}) and (\ref{SE}). The discrepancy between approaching $x=0$ from two sides is given by $J^2/(E-h_0)$ with $h_0$ being the constant term in $h(\beta)$, as $n = 0$.

\subsection{Scattering wavefunction}
Having in mind the above discussions, let us write down $\psi(x)$ in Eq.~(\ref{pxL}) for a scattering state in a more explicit form in terms of the (generalized) self-energy in the thermodynamic limit. As any eigenstate with eigen-energy apart from the band dispersion is necessarily a bound state \cite{GongPRA2022}, a scattering state always has an eigen-energy $E$ (almost) on the band dispersion for finite systems (though the converse is not always true, as there could be bound states in the continuum \cite{hsu2016bound}). It is thus natural to write $E=h_{\tilde k} =h(e^{i\tilde k})$ with $\tilde k$ almost real, i.e., the real momentum is analytically continued into the complex plane to capture the finite-size discrepancy between $E$ and the dispersion. For simplicity, as is also typically the case, we assume other solutions $y$ (other than $e^{i\tilde k}$) to $E=h(\beta)$ exhibit norms well deviated from $1$. 

We first recall that $E=h_{\tilde k}$ satisfies $E-\Delta - \Sigma^{(L)}(E)=0$, which is well approximated by
\begin{equation}
h_{\tilde k} - \Delta - \Sigma^{<}(h_{\tilde k}) - \frac{J^2}{ih'_{\tilde k}(1-e^{i\tilde k L})} \simeq 0,
\label{ex}
\end{equation} 
where we have used $h'_k= ih'(e^{ik})e^{ik}$ (chain rule), $\Sigma^{<}$ refers to the branch of $\Sigma$ (thermodynamic-limit self-energy) whose the first (second) line of residue formula in Eq.~(\ref{SE}) excludes (includes) $e^{i\tilde k}$, and $\simeq$ means up to exponentially small error in terms of $L$. The other branch is related to the above one via
\begin{equation}
\Sigma^{>}(h_{\tilde k}) - \Sigma^{<}(h_{\tilde k}) = \frac{J^2}{i h'_{\tilde k}},
\label{Sd}
\end{equation}
(also see Appendix \ref{app:discont-self-energy}) 
so we also have
\begin{equation}
h_{\tilde k} - \Delta - \Sigma^{>}(h_{\tilde k}) - \frac{J^2e^{i\tilde k L}}{ih'_{\tilde k}(1-e^{i\tilde k L})} \simeq 0.
\label{in}
\end{equation}
Combining Eqs.~(\ref{ex}) and (\ref{in}), we obtain
\begin{equation}
e^{i\tilde k L} \simeq \frac{h_{\tilde k} - \Delta - \Sigma^{>}(h_{\tilde k})}{h_{\tilde k} - \Delta - \Sigma^{<}(h_{\tilde k})}= \frac{G_e^<(h_{\tilde k})}{G_e^>(h_{\tilde k})},
\label{eq:ratio}
\end{equation}
where we have introduced the emitter Green's function associated with each branch:
\begin{equation}
G_e^{>,<}(z)= \frac{1}{E-\Delta-\Sigma^{>,<}(z)}.
\end{equation}

Suppose $\tilde k$ is perturbed from a real $k$. Then the shift should be a series of $L^{-1}$, and in particular the leading order of the imaginary part is determined by
\begin{equation}
{\rm Im}\tilde k = \frac{1}{L}\log\left | \frac{G^{>}_e(h_{k})}{G^{<}_e(h_{k})} \right | 
+ \mathcal{O}(L^{-2}).
\label{eq:imk}
\end{equation}
Here we have assumed $G_e^{>,<}(z)$ is smooth near $h_k$, so that the shift from $k$ to $\tilde k$ only causes higher-order corrections.

We move on to the corresponding eigenstate. Any exact photon wave function follows from substituting Eq.~(\ref{SR}) into Eq.~(\ref{pxL}). As for a scattering state with energy $E=h_{\tilde k}$, the wave function can be well-approximated by 
\begin{equation}
\psi_{\rm s}(x)\simeq \left\{
\begin{array}{ll} 
\frac{J }{ih'_{\tilde k} (1- e^{i\tilde k L})}e^{i\tilde k x} + J^{-1}\Sigma^{<}_x(h_{\tilde k}) & x\ge0\\
\frac{J }{ih'_{\tilde k} (e^{-i\tilde k L} - 1)}e^{i\tilde k x} + J^{-1} \Sigma^{>}_x(h_{\tilde k})  & x<0.
\end{array}\right.
\label{ps}
\end{equation}
Here we have omitted the unimportant normalization factor. Just like $\Sigma^{< (>)}$, $\Sigma^{< (>)}_x$ refers to the first (second) line of the residue formula in Eq.~(\ref{Sx}) with $y=e^{i\tilde k}$ excluded. As a natural generalization of Eq.~(\ref{Sd}), the difference from the other branch is given by
\begin{equation}
\Sigma^{>}_x(h_{\tilde k}) - \Sigma^{<}_x(h_{\tilde k}) =\frac{J^2}{ih'_{\tilde k}}e^{i\tilde kx}. 
\end{equation}
Recalling Eqs.~(\ref{ex}) and (\ref{in}), we find
\begin{align}
    \begin{split}
       \psi_{\rm s}(x)&\simeq J^{-1}[G_e^<(h_{\tilde k})^{-1} e^{i\tilde k x} + \Sigma^<_x(h_{\tilde k})] \\
    &= J^{-1}[G_e^>(h_{\tilde k})^{-1} e^{i\tilde k x} + \Sigma^>_x(h_{\tilde k})].
    \end{split}
\label{wavefunction_finiteL_branch}
\end{align}
It turns out that the choice of branch does not really matter, so we can safely drop the superscript $<,>$ in the formal discussion. 

Also, by multiplying a proper coefficient, we indeed end up with the LS wave function \cite{Lippmann1950} (see Eq.~(\ref{eq:LS-wavefunction}) and Appendix \ref{app:lattice-LS}):
\begin{equation}
\psi_{k}(x) = e^{i\tilde k x} + G_e(h_{\tilde k}) \Sigma_x (h_{\tilde k}).
\label{eq:wavefunction_finiteL}
\end{equation}

At first glance, it may be rather surprising that the choice of branch is irrelevant since it is well-known that the choice does matter in continuous Hermitian systems, and it corresponds to either forward or backward scattering process \cite{CohenAP}. More explicitly, the two branches of the self-energy are related by complex conjugation, and likewise one branch of $\Sigma_x$ is related to the other branch of $\Sigma_{-x}$ by complex conjugation. 

However, there is no contradiction, as Hermitian systems are atypical in the NH world. More precisely, a Hermitian band can be considered as consisting of self-intersecting points (almost) everywhere, while self-intersecting points are rare in NH systems. As will become clear in the following, the above analysis for typical cases (i.e., points lying on a boundary separating \emph{two} regions with different spectral winding numbers; in contrast, a self-intersecting point should be a ``multi-critical point'' of at least three regions) does not apply to self-intersecting points or occasional degeneracy. It is thus natural to expect a total breakdown for Hermitian systems. In addition, the explicit meanings of branches are actually different. In the NH case, different branches mean analytical continuations from different regions separated by the band dispersions. In the Hermitian case, we are actually talking about the \emph{same} branch (outside the collapsed loops) in the NH sense, while different branches here arise from intrinsic singularities (i.e., branch points). Therefore, a fair comparison should be made for the specific NH branch outside the loops if one tries to take the Hermitian limit.

\subsection{Other circumstances}
\label{sub:other}
In the previous section, we have developed a general approach to typical eigenvectors of Eq.~(\ref{eq:Eigenvalue-conditions}) under physical assumptions, which is expected to be valid for generic NH systems. In this section, we discuss three atypical cases where the exact eigenvectors are not in the form of Eq.~(\ref{eq:wavefunction_finiteL}), the last of which is the well-known bound states. 

\subsubsection{Spectral degeneracy}
\label{sub:spectral-deg}
The LS wavefunctions in Eq.~(\ref{eq:wavefunction_finiteL}) are obtained under the assumption of $c_e \neq 0$. This is a highly physical assumption as the emitter-photon interaction is finite. Here we discuss the cases where this assumption is violated. 

For $c_e = 0$, the exact eigenvalue equation Eq.~(\ref{eq:Eigenvalue-conditions}) reduces to 
\begin{align}
    \begin{split}
        \sum_k c_k &=0, \\
    (E-h_k)c_k &=0 , \forall  k=2\pi m/L.
    \end{split}
\end{align}
Accordingly, the scattering wavefunctions are given by a linear superposition of plane waves. This can happen in the cases of spectral degeneracy. Then for each degenerate $h_k$, which means there are other $l\in\mathbb{Z}^+$ different $k'$ (from each other and from $k$) such that $h_{k'}=h_k$, we have $l$ linearly independent degenerate solutions with $E=h_k$ and $\sum_{k': h_{k'}=h_k} c_{k'}=0$.

We emphasize this situation should be atypical. That is, without further assumptions such as Hermiticity or symmetry, degeneracy only occurs occasionally. Moreover, even if there is a degeneracy in the thermodynamic limit where $k$ takes on a continuous value over $(0,2\pi]$, the degeneracy may no longer be exact in the finite size case, as $k=2\pi m/L$ only takes on finite discrete values. A similar finite-size effect also appears in $c_e$, such that $c_e$ is (almost) never strictly zero.

Clearly, the occurrence of such fine-tuned situations implies NH dispersions with self-intersecting points. Despite that $h_k$ should be non-degenerate for almost all $k$, the existence of self-intersecting points can be enforced by NH band topology. Consider (single-band) NH baths with complex dispersion relations $h_k$ with non-zero winding numbers defined by \cite{Gong2018}
\begin{equation}
    \text{ind}( h_k -z)  =  \int^{\pi}_{-\pi} \frac{dk}{2 \pi i} \partial_k \ln \det(h_k - z ). 
\label{eq:winding-number}
\end{equation}
For a $h_k$ with a maximum winding number $|w|$, there are at least $|w|-1$ points (including multiplicity) in the complex plane where $h_k$ self-intersects. This follows from an old result proved by Whitney \cite{Whitney1937}. One way of understanding this fact is that the winding number can only change $\pm$ 1 when $z$ crosses any loop defined by $h_k$ on the complex plane, and the sign depends on the orientation of the $h_k$ loop. Without self-intersection, any two locally parallel parts of the loop are oriented in the reversed direction. For the same $h_k$, there are also points $z'$ in regions that have a winding number of zero. The straight line that connects $z$ from the maximum winding region to $z'$ always exists, and it must cut $h_k$ with the correct orientation. %And 
Intuitively, this can only happen if there must be at least $|w|-1$ self-intersecting points, since $h_k$ is a continuous map from a circle and different loop components are necessarily connected somewhere. 
Note that since we focus on single-band baths, the self-intersecting points are normal spectral degeneracies and not exceptional points. This feature is illustrated analytically in a minimal model in Sec.~\ref{sec:NNN}. 

For such atypical situations, the scattering wavefunction is no longer given by Eq.~(\ref{eq:wavefunction_finiteL}). Because for generic situations, only one root $E = h(y)$ has a unit magnitude. But for fine-tuned situations, there can be multiple unit magnitude roots $e^{i\tilde k_s}$ with $s=1,2,...$. 
The corresponding finite-size terms $e^{i\tilde{k}_s L}$ do not tend to zero or infinity in the thermodynamic limit. 
Therefore, the exact eigenvalue equation $E-\Delta - \Sigma^{(L)}(E)=0$ is no longer approximated by Eq. ($\ref{ex}$), but rather
\begin{equation}
h_{\tilde k} - \Delta -  \Sigma^{<}(h_{\tilde k}) - \sum_{s} \frac{J^2}{ih'_{\tilde k_s}(1-e^{i\tilde k_s L})}  \simeq 0.
\label{ex_fine_general}
\end{equation}

\subsubsection{Root multiplicity}
\label{sec:root-multi}

There is another situation for the LS wavefunctions to break down. In the derivation of Eq.~(\ref{eq:wavefunction_finiteL}), we have assumed all the roots of $E = h(y)$ do not have a multiplicity. This assumption is valid for generic situations. But this assumption can be violated in fine-tuned situations. The roots of $E = h(y)$ can be fine-tuned to have a multiplicity and the evaluations of the residue formula Eq. (\ref{xL}) require a higher-order pole formula. 

In general, the root multiplicity of $E-h(y)=0$ has an interesting connection with the derivatives of $h_k$. The dispersion $h_k$ has vanishing derivatives at orders $r-1, r-2 , ..., 1$ if and only if the root multiplicity is $r$. This can be seen by rewriting the root condition as $y^q(h(y)-E)/h_{-p} = \prod_n (y-y_n)^{r_n} $ and evaluating the $r$-th order derivatives at $y = y_n$, which is equivalent to calculating the same derivatives at the corresponding $y_n$.

This occurs even in Hermitian systems, such as the band edges. Therefore, one expects the LS wavefunction to have an alternative expression. However, an explicit calculation of such scattering states for the NNN baths using the higher-order pole formula does not match the exact diagonalization results for finite systems. This is because the root multiplicity is never exact in finite systems and the poles are again simple. Recall that the scattering wave functions are extremely sensitive to the tiny shift in the wave numbers; it is not surprising that a na\"ive application of the higher-order pole formula, which essentially concerns the thermodynamic limit, does not work.

Here we focus on the cases of degenerate unit magnitude roots, which appear in the case of the NNN baths. 
Without loss of generality, the exact eigenvalue equation $E-\Delta - \Sigma^{(L)}(E)=0$ is approximated by 
\begin{equation}
h_{\tilde k} - \Delta -  \Sigma^{<}(h_{\tilde k}) - \sum_{r} \frac{J^2}{ih'_{\tilde k_r}(1-e^{i\tilde k_r L})}  \simeq 0,
\label{ex_roots_general}
\end{equation}
with $r$ being the root multiplicity, and the root multiplicity is (almost) never exact for any finite systems. The first three terms are $\mathcal{O}(1)$, so the remaining sum of degenerate root terms must also be $\mathcal{O}(1)$ to make the approximated equation hold.

Moreover, having a root multiplicity ($r \neq 1$) is a fine-tuned case for NH systems, and the cases $r>2$ are even more atypical. Thus, we consider the case of $r=2$, where the finite-size expression Eq. (\ref{ex_roots_general}) is well-approximated by 
\begin{equation}
h_{\tilde k} - \Delta -  \Sigma^{<}(h_{\tilde k})- \frac{J^2}{ih'_{\tilde k_\alpha}(1-e^{i\tilde k_\alpha L})} - \frac{J^2}{ih'_{\tilde k_\gamma}(1-e^{i\tilde k_\gamma L})} \simeq 0,
\label{ex_fine}
\end{equation} 
where $\tilde k_{\alpha,\gamma}$ are the two corresponding finite-size momenta. 

The analysis above indicates $\tilde k_{\alpha, \gamma}$ and $h'_{\tilde k_\alpha, \gamma}$ must adopt a simple form. To see this explicitly, consider the following finite size corrections to $k_r$, $\tilde k_{\alpha,\gamma}=k_r\pm c_r L^{-1}+\mathcal{O}(L^{-2})$, where $c_r$ is an $\mathcal{O}(1)$ coefficient. Here we have used the fact that $\tilde k_{\alpha,\gamma}$ are solutions to $E = h_{\tilde k}$ with a fixed $E$ that is close to $h_{k_r}$. Recall that the sum of the first three terms remains $\mathcal{O}(1)$. For Eq.~(\ref{ex_fine}) to be valid,  
we conclude that the remaining two $\mathcal{O}(L)$ finite size terms, a point that will be justified later, should cancel out at the leading order. This leads to $e^{i\tilde k_\alpha L}=e^{i \tilde k_\gamma L}$ and thus $e^{2ic_r}=1$. Overall, we rewrite the finite size corretions as $\tilde k_{\alpha, \gamma} = k_r \pm m \pi L^{-1} + \mathcal{O}(L^{-2})$ where $m$ is a positive integer. Finally let us return to the fact that each of $h'_{\tilde k_\alpha, \gamma}$ is $\mathcal{O}(L^{-1})$ and $J^2/\left( ih'_{\tilde k_{\alpha,\gamma}}(1-e^{i\tilde k_\gamma L}) \right)$ is $\mathcal{O}(L)$. This fact follows from examining the Taylor expansion of the first-order derivative around $k_r$, $h'_{\tilde k_{\alpha, \gamma} } = h'_{ k_{r} } \pm m \pi h^{''}_{ k_{r}} L^{-1} + \mathcal{O}(L^{-2})$. Recall that $h'_{ k_{r} }=0$, so the non-vanishing leading term is the second-order derivative.

\subsubsection{Bound States}
\label{subsec:bound-states}
The exact eigenvalue equation Eq.~(\ref{eq:Eigenvalue-conditions}) contains the spatially delocalized scattering states as well as bound states. A bound state has a spatially localized particle density profile around the emitter and has a non-vanishing $c_e \neq 0$ in the thermodynamic limit. Bound states in NH systems have been studied in Refs.~\cite{GongPRL2022,GongPRA2022} and their impact on photon-mediated interactions are addressed in Refs.~\cite{RoccatiOptica2022,Sun2025}. Here we revisit this topic and provide some explicit results.     
    
For a bound state with an exact energy $E_{b}$, eliminating $c_e$ in Eq.~(\ref{eq:Eigenvalue-conditions}) gives Eq. (\ref{eq:Eigenvalue-conditions_ceneq0}) evaluated at $E = E_b$, whose thermodynamic limit reads
\begin{equation}
    E_b - \Delta - \Sigma(E_b) = 0. 
\end{equation}
Let us recall that the bound state energy is well-separated from the band dispersion and one can take the thermodynamics limit of its finite-size self-energy without technical difficulties.

As shown in Ref. \cite{GongPRL2022}, there are also anomalous bound states as a unique feature of NH systems exhibiting skin effects. The poles of the emitter Green's functions $G_e(z)$ contain information about all the bound states and predict the corresponding $E_{b}$ exactly \cite{GongPRA2022}. This will be shown explicitly for the Hatano-Nelson and unidirectional NNN baths in Sec. \ref{sec:HN} and Sec. \ref{sec:NNN}.

Let us comment a bit on the relation between bound states and scattering. In standard scattering theories for continuous Hermitian systems, bound states are poles of the $S$-matrix, which is closely related to the poles of the reflection and transmission coefficients. However, this calculation is tricky to perform for NH baths because the $S$-matrix is generally ill-defined as the quantum dynamics is intrinsically non-unitary. It may be interesting to explore to what extent one can extend the notion of $S$-matrix and related concepts to NH systems.

\section{Hatano-Nelson Baths}
\label{sec:HN}
\begin{figure}[t]
  \centering
  \includegraphics[width=1\columnwidth]{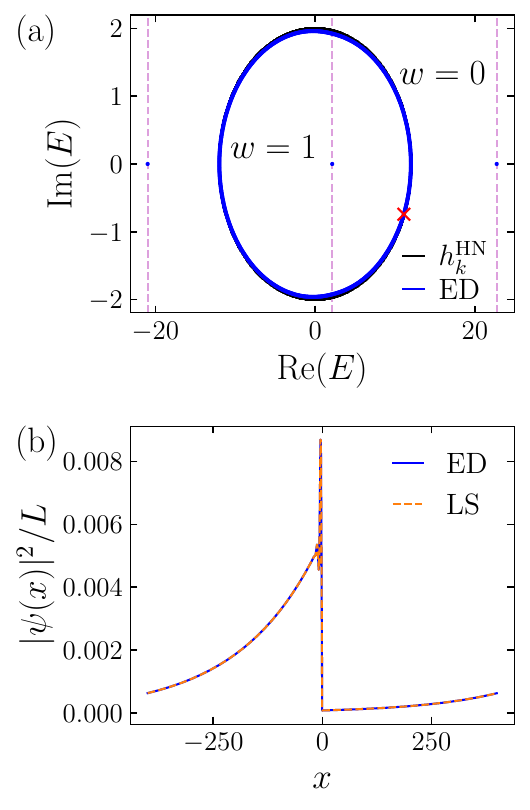}%{Fig2.pdf}
  \caption{ HN baths for $u=6,\ \kappa=2,\ J=20,\ \Delta=2.14$ and $L=801$. (a) The single-excitation spectrum of the $H$ obtained from exact diagonalization is shown in blue. The black dotted line indicates the dispersion $h^{\text{NH}}_{k}$. The pink dashed vertical lines indicate all the bounds states, and the positions are analytically predicted by the relevant emitter Green's functions (see main text for the numerical values). An eigenvector is selected at random (red marker) and shown in (b) (blue), which shows a good agreement with the analytical LS wave function under PBC (see main text for the discussion). We have normalized the LS wavefunction Eq.~(\ref{eq:LS-wavefunction-HN}) by dividing $\sqrt{L}$, such that Eq.~(\ref{eq:Normalization}) holds.
  }  
  \label{fig:fig2}
\end{figure}

\begin{figure}[t]
  \centering
  \includegraphics[width=1\columnwidth]{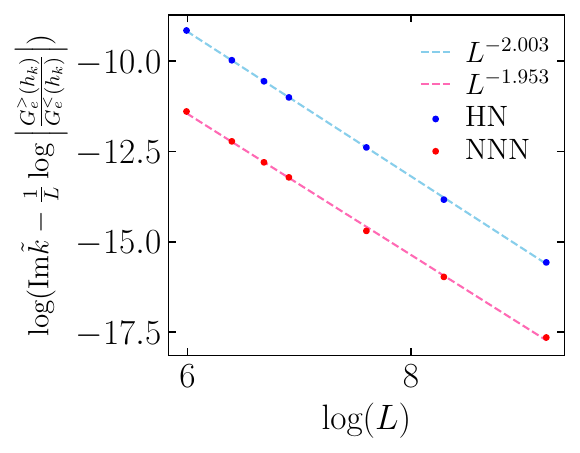}
  \caption{For each system sizes, the value of ${\rm Im}\tilde k$ changes for a fixed scattering states with an exact eigenvalue $E$. The difference between ${\rm Im}\tilde k$ and its leading order in $L^{-1}$ shows a $\mathcal{O}(L^{-2})$ scaling for both NN and NNN baths, as predicted by Eq. (\ref{eq:imk}).} 
  \label{fig:fig5}
\end{figure}

In this section, we consider the NH baths to be the Hatano-Nelson (HN) model as a concrete example \cite{Hatano1996}. The full derivation of the result is given in Appendix \ref{app:derivation-HN}. The HN model is the simplest NH generalization of the tight-binding model, where only the nearest-neighbor is included and the left and right hopping amplitudes are not the same. The NH bath Hamiltonian Eq.~(\ref{eq:Ham-realspace}) reads $h_{x-1,x} = -(u - \frac{\kappa}{2})$, $h_{x+1,x} = -(u + \frac{\kappa}{2})$ and zero otherwise. Such an effective NH bath can be realized by choosing non-local Lindblad operators \cite{Metelmann2015}. 

According to Eq.~(\ref{eq:hk}), the band dispersion is

\begin{equation}
    h^{\text{HN}}_k = -\left(u-\frac{\kappa}{2}\right)e^{ik} -\left(u+\frac{\kappa}{2}\right)e^{-ik}.
\label{eq:hk-HN}
\end{equation}
It has a maximum winding number $|w| = 1$ defined by Eq.~(\ref{eq:winding-number}) and does not contain any self-intersecting points. The spectrum of the full Hamiltonian Eq.~(\ref{eq:NH-Ham-k}) is shown in Fig.~\ref{fig:fig2} and it is similar to the structure of $h^{\text{HN}}_k$ except for three discrete points %\textcolor{red}{[ZG: Better to make the points larger and above the dashed lines.]} \jl{I will fix this in the end}. 
These points correspond to the bound states, to which we refer later. 

For finite systems, the LS scattering wavefunction Eq. (\ref{eq:wavefunction_finiteL}) for the HN baths are the following:
\begin{widetext}
\begin{align}
\begin{split}
&\psi^{>}_{\tilde k}(x) =
    \begin{cases}
 e^{i\tilde{k}x} + G^{>}_e( h^{\text{HN}}_{\tilde k}  )(\frac{J^2}{\delta})e^{i\tilde{k}x} & x\ge0 \\
      e^{i\tilde{k}x} + G^{>}_e( h^{\text{HN}}_{\tilde k}  )\left(\frac{J^2}{\delta}\left(\frac{u-\frac{\kappa}{2}}{u+\frac{\kappa}{2}}\right)^{-x}\right)e^{-i\tilde{k}x} & x<0,
    \end{cases}     
    \end{split}
\label{eq:LS-wavefunction-HN}
\end{align}
\end{widetext}
where $\delta = (u - \kappa/2)e^{i\tilde{k}} - (u +\kappa/2)e^{-i \tilde{k}}$ and 
\begin{align}
\begin{split}
G^{>}_e( h^{\text{HN}}_{\tilde k}  ) &= \frac{1}{h^{\text{HN}}_{\tilde{k}} -\Delta-\frac{J^2}{\delta}}
\end{split}
\end{align}
is the emitter Green's function. The analytical form is obtained by calculating all the relevant Green's functions in Eq.~(\ref{TL}). 

The finite-size expression $\psi^{>}_{\tilde k}(x)$ approaches to the $>$ branch with scattering wavefunction $\psi^{>}_{k}(x)$ in the thermoydnamics limit. The predicted finite-size scaling of ${\rm Im}\tilde k$ at the leading order is numerically verified in Fig.~\ref{fig:fig5}.

Equivalently, one can also consider the other thermodynamic branch $\psi^{<}_{k}(x)$, with its finite-size expression $\psi^{<}_{\tilde k}(x)$ given by Eq.~(\ref{eq:LS-wavefunction-NH-unphysical}). At first glance, $\psi^{>}_{\tilde k}(x)$ and $\psi^{<}_{\tilde k}(x)$ are not identical. But in fact, the two branches are related via Eq.~(\ref{wavefunction_finiteL_branch}) exactly.
In Fig.~\ref{fig:fig2} (b), we show a good agreement between LS scattering wavefunction evaluated at $\tilde k$ that solves the exact eigenvalue equation Eq.~(\ref{eq:Eigenvalue-conditions_ceneq0}) and the exact diagonalization result.

The real-space behavior of the NH scattering states is qualitatively different from the usual Hermitian story. In general, all the NH scattering states exhibit a large localization length that is proportional to the system size. Fig.~\ref{fig:fig2} shows the exact diagonalization results for PBC, and (b) illustrates the spatial amplitude of one eigenvector chosen randomly (all the other eigenvectors have a similar spatial property). The exact scattering eigenvectors have an inhomogeneous amplitude in real space. This is in contrast to the usual plane wave solutions, where the spatial amplitudes are uniform.

In the single-excitation sector, not all the exact eigenvectors are scattering states. As shown in Fig.~\ref{fig:fig2}, there are three eigenvalues far away from $h_k$ and their spatial density is localized around the emitter $x=0$, known as the bound states. Structured NH baths can induce hidden bound states in addition to the usual bound states in Hermitian systems. The hidden bound state is located in the winding number $|w|=1$ domain, and the usual bound states are in the trivial winding number domain. As mentioned in Sec.~\ref{subsec:bound-states}, the poles of emitter Green's function (\ref{eq:emitter-Green-HN}) predict the energies $E$ of all the bound states. In turn, the profile and localization length can be obtained from Eq.~(\ref{pxL}), so formally a bound state can be considered as a scattering state with complex wave numbers $\tilde k$ (whose imaginary parts are $\mathcal{O}(1)$) such that $E=h_{\tilde k}$. The hidden bound states correspond to imaginary momentum that solves $1/G^{<}_e( h^{\text{HN}}_{\tilde k} ) = 0$, and we find the numerical value to be $\tilde{k} \approx 1.175 -0.168i$ in this case. In addition, the two usual bound states can be obtained by solving $1/G^{>}_e( h^{\text{HN}}_{\tilde k}) = 0$, which gives $\tilde{k} \approx 1.006i $ and $\tilde{k} \approx \pi + 1.102i$.

To avoid potential confusion, we stress that the results of this section are calculated under the assumption that $h^{\text{HN}}_{k}$ does not exhibit spectral degeneracy, which is true for $\kappa \neq 0$. In the Hermitian limit ($\kappa = 0$), the HN baths reduce to the nearest-neighbor tight-binding model, and the dispersion is given by $-2u\cos(k)$. Due to the degeneracy for each pair of $k$ and $-k$, the scattering states for finite systems are not to be understood as the $\kappa \rightarrow 0$ limit of Eq. (\ref{eq:LS-wavefunction-HN}). Instead, one has to take the degeneracy for each $k$ into account when approximating the exact eigenvalue equation $E-\Delta - \Sigma^{(L)}(E)=0$. We hope to report some concrete connections to Hermitian systems in future studies.

\section{unidirectional next-to-nearest-neighbor Baths}
\label{sec:NNN}
\begin{figure}[t]
\centering
\includegraphics[width=1\columnwidth]{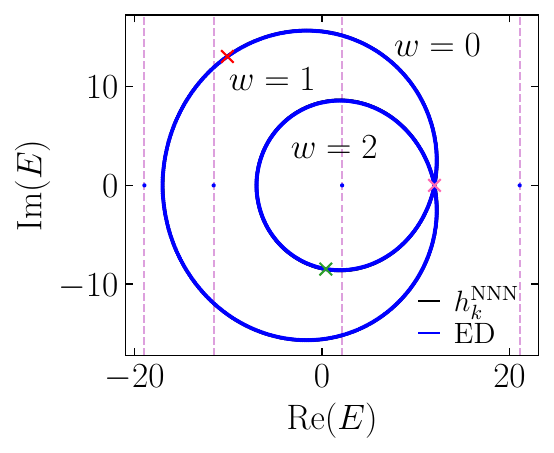}
\caption{NNN baths for $\kappa=5, \kappa'=12, L=801, J=20, \Delta=2.14$. The single-excitation spectrum of the $H$ obtained from exact diagonalization is shown in blue. The black dotted line indicates the dispersion $h^{\text{NNN}}_{k}$. The pink dashed vertical lines indicate all the bound states. Two ED eigenvectors in $k_{1}$ and $k_{2}$ are chosen at random, indicated by the red and green markers. The pink marker indicates the self-intersecting state. The real space distributions of states indicated by the red, green, and pink markers are shown in Fig. \ref{fig:fig3_2}.}  
\label{fig:fig3_1}
\end{figure}

\begin{figure}[t]
\centering
\includegraphics[width=0.95\columnwidth]{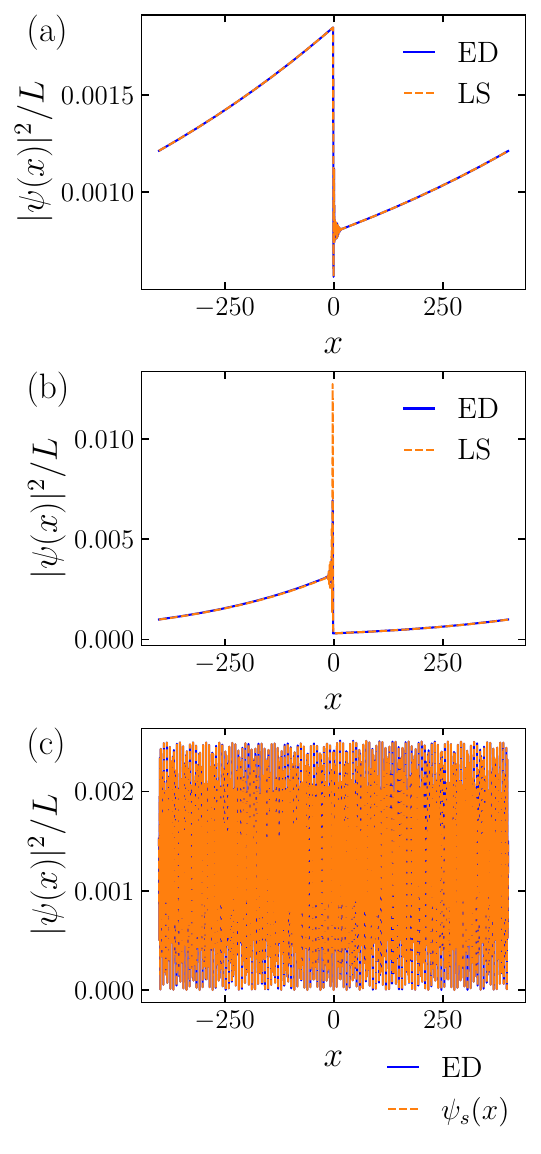}
\caption{NNN baths for $\kappa=5, \kappa'=12, L=801, J=20, \Delta=2.14$. The states indicated by the red, green, and pink markers in Fig. \ref{fig:fig3_1} are shown in (a), (b), and (c), respectively, and compared with the analytical LS expressions under PBC. The LS wavefunctions are normalized according to Eq.~(\ref{eq:Normalization}).}  
\label{fig:fig3_2}
\end{figure}

Having studied the HN baths, we next consider a more complicated model, unidirectional next-to-nearest-neighbor baths. The asymmetric hopping range is extended to $2$ but restricted to a single direction. The NH baths Hamiltonian Eq.~(\ref{eq:Ham-realspace}) reads $h_{x+1,x} = -\kappa$, $h_{x+2,x} = -\kappa'$ and zero otherwise. The complex dispersion reads
\begin{equation}
    h^{\text{NNN}}_k = -\kappa e^{-ik} - \kappa'e^{-2ik}.
\label{eq:hk-NNN}
\end{equation}
The NNN dispersion $h^{\text{NNN}}_k$ has a maximum winding number 2 defined by Eq.~(\ref{eq:winding-number}). This makes finding the explicit form of all the Green's functions (cf. Eq.~(\ref{eq:Green-self-energy})) difficult, as will be explained in this section. There is a single point in the complex plane where $h^{\text{NNN}}_k$ with two different momenta self-intersect. This happens for states with momenta 
\begin{equation}
    k_{\text{SI}}=\arccos{\left(-\frac{\kappa}{2\kappa'}\right)}
\end{equation}
with an energy $h^{\text{NNN}}_{k_{\text{SI}}}= \kappa'$. Furthermore, $h^{\text{NNN}}_k$ separates three domains of different winding numbers. The outer boundary of the winding number $1$ domain (the ``outer-loop'' in Fig.~\ref{fig:fig3_1}), is given by $h^{\text{NNN}}_{k_1}$, where $|k_1| %= |k| 
> | k_{\text{SI}}|$. 
In addition, $h^{\text{NNN}}_{k_2}$ for $ |k_2| %= |k| 
< | k_{\text{SI}}|$ 
defines the boundary of the winding number $2$ domain (the ``inner-loop''). 

For a finite number of lattice sites, the explicit forms of LS wavefunction (cf. Eq.~(\ref{eq:wavefunction_finiteL})) are 
\begin{widetext}
\begin{align}
    \begin{split}
        \psi^{>}_{\tilde k_1}(x) = 
    \begin{cases}
        e^{i\tilde{k}_1x}+G^{>,x \ge 0}_e( h^{\text{NNN}}_{ \tilde{k}_{1}}   )\frac{J^2}{\delta} \left((-\frac{\kappa'}{\kappa+\kappa'e^{-i\tilde{k}_1}})^{x+1}-e^{i\tilde{k}_1(x+1)} \right) & x \ge 0 \\
        e^{i\tilde{k}_1x} & x<0,
    \end{cases} 
    \label{eq:LS-wavefunction-NNN1}
    \end{split}\\
    \begin{split}
        \psi^{>}_{\tilde k_2}(x) = 
    \begin{cases}
        e^{i\tilde{k}_2x}-G^{>,x\ge0}_e( h^{\text{NNN}}_{\tilde{k}_{2}}  )\frac{J^2}{\delta}e^{i \tilde{k}_2(x+1)} & x \ge 0 \\
        e^{i\tilde{k}_2x}-G^{>,x<0}_e( h^{\text{NNN}}_{\tilde{k}_{2}}   )\frac{J^2}{\delta}(-\frac{\kappa'}{\kappa+\kappa'e^{-i\tilde{k}_2}})^{x+1} & x<0,
    \end{cases}
    \label{eq:LS-wavefunction-NNN2}
    \end{split}
\end{align}
\end{widetext}
where $\delta=\kappa+2\kappa'e^{-i\tilde{k}_{1/2}}$ depending on the value of $\tilde{k}$, and 
\begin{align}
\begin{split}
G^{>,x\ge 0}_e( h^{\text{NNN}}_{\tilde k_{1}}  ) &=
\frac{1}{h^{\text{NNN}}_{k_1} -\Delta + J^2(\frac{e^{ik_1}}{\kappa+\kappa'e^{-i\tilde{k}_1}})},  \\
G^{>,x\ge0}_e(h^{\text{NNN}}_{\tilde k_{2}}  ) &= \frac{1}{h^{\text{NNN}}_{\tilde{k}_2} -\Delta  +\frac{J^2}{\delta}e^{i\tilde{k}_2}},  \\
G^{>,x < 0}_e( h^{\text{NNN}}_{\tilde k_{2}}   ) &= \frac{1}{h^{\text{NNN}}_{\tilde{k}_2} -\Delta -\frac{J^2}{\delta}(\frac{\kappa'}{\kappa+\kappa'e^{-i\tilde{k}_2}})}
\end{split}
\end{align}
are the emitter Green's function evaluated for the corresponding arguments and the full derivation is given in Appendix \ref{app:derivation-NNN}. The two sets of states are defined on $k_1$ and $k_2$, separated by the self-intersecting point $k_{\text{SI}}$. And the excellent agreement with the ED results is displayed in Fig. \ref{fig:fig3_2} (a) and (b). The self-intersecting point in (c) is discussed later. In the thermoydnamic limit, $\psi^{>}_{\tilde k_{1/2}}(x)$ approaches the corresponding branch $\psi^{>}_{k_{1/2}}(x)$. The predicted finite-size scaling of ${\rm Im}\tilde k$ at the leading order is numerically verified in Fig.~\ref{fig:fig5}. 

Equivalently, one could consider $\psi^{<}_{\tilde k_{1/2}}(x)$, which corresponds to another thermodynamic branch $\psi^{<}_{k_{1/2}}(x)$. The explicit form is given by Eq.~(\ref{eq:LS-wavefunction-NNNother}).  Again, the two branches are related via carefully using the corresponding Green's function Eq.~(\ref{eq:emitter-Green-NNN}). 

Lastly, the four bound states are identified with their corresponding emitter Green's functions, see Fig.~\ref{fig:fig3_1}. We numerically find that the complex momenta that solves $1/G^{>,x\ge 0}_e( h^{\text{NNN}}_{\tilde k_{1}}  ) = 0$ produce the usual bound states, $\tilde{k} \approx0.064i$ and $\tilde{k} \approx 1.729+0.282i$. Similarly, the hidden bound states in winding number 1 and 2 domains are given by poles $G^{<,x\ge0}_e(h^{\text{NNN}}_{\tilde k_{1}}  )$ (with $k\approx -0.229i$)  and $G^{>,x\ge 0}_e( h^{\text{NNN}}_{\tilde k_{2}} )$ (with $\tilde{k} \approx 2.087-0.862i$). 

\section{Fine-tuned cases}
\label{sec:fine-tuned-points}
So far, we have only considered the scattering states that are described by the LS wavefunction given by Eq.~(\ref{eq:wavefunction_finiteL}). The validation is confirmed by benchmarking our analytical results with the exact diagonalization for finite lattices, which shows an excellent agreement. This LS wavefunction captures scattering states for generic hopping parameters under the mild assumption of a nondegenerate spectrum and simple poles. However, as mentioned in Sec.~\ref{sub:other}, there are special fine-tuned scenarios when the assumption breaks. 

In this section, we provide minimal examples that appeared in the NNN models, where we have a total hopping range of $2$, spectral degeneracy occurs when the two simple poles $e^{i\tilde k_\alpha}$ and $e^{i\tilde k_\gamma}$ of $(E-h(y))^{-1}$ have an almost unit magnitude for finite systems. In this case, the finite-size expression $E-\Delta - \Sigma^{(L)}(E)=0$ is well-approximated by Eq. (\ref{ex_fine}) and the scattering wavefunctions are given by 
\begin{equation}
\psi_{\rm s}(x)\simeq \left\{
\begin{array}{ll} 
\frac{J }{ih'_{\tilde k_\alpha} (1- e^{i\tilde k_\alpha L})}e^{i\tilde k_\alpha x} + \frac{J }{ih'_{\tilde k_\gamma} (1- e^{i\tilde k_\gamma L})}e^{i\tilde k_\gamma x}  & x\ge0 \\
\frac{J }{ih'_{\tilde k_\alpha} (e^{-i\tilde k_\alpha L} - 1)}e^{i\tilde k_\alpha x} + \frac{J }{ih'_{\tilde k_\gamma} (e^{-i\tilde k_\gamma L} - 1)}e^{i\tilde k_\gamma x}  & x<0.
\end{array}\right.
\label{ps_fine}
\end{equation}

\subsection{Self-intersecting point}
\begin{figure}[t]
\centering
\includegraphics[width=1\columnwidth]{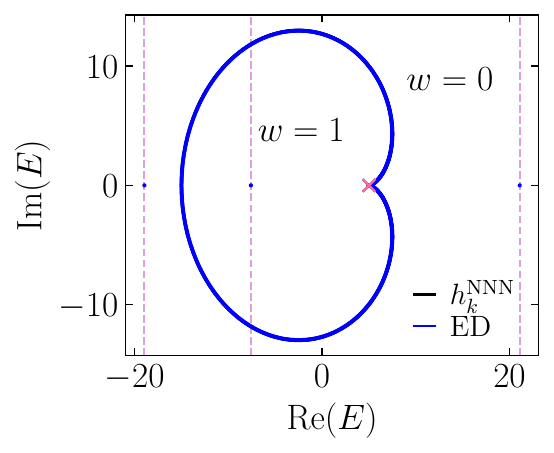}
\caption{NNN baths for $\kappa= 10, \kappa'= 5, L= 801, J=20, \Delta=2.14$. The single-excitation spectrum of the $H$ obtained from exact diagonalization is shown in blue. The black dotted line indicates the dispersion $h^{\text{NNN}}_{k}$. The pink dashed vertical lines indicate all the bound states. Under the fine-tuned parameter $\kappa = 2\kappa' $, the winding numbers are only $0$ and $1$. The cross indicates the fine-tuned state. }
\label{fig:fig4_1}
\end{figure}
A special case occurs when the $h^{\text{NNN}}_{k}$ dispersion has a maximum winding number of 2, which means there is a self-intersecting point for generic parameters. In the thermodynamic limit, at the self-intersecting point, there are two simple poles, $e^{ik_{\text{SI}}}$ and $e^{-ik_{\text{SI}}}$ (see Appendix \ref{app:derivation-NNN}). For any finite systems, $c_e$ is almost never strictly zero and $\tilde{k}_{\text{SI}}$ turns out to have an almost vanishing (at least smaller than $\mathcal{O}(L^{-1})$) imaginary part. 
Evaluating the scattering wavefunction for $\tilde{k}_{\text{SI}}$, which shows a perfect matching in Fig.~\ref{fig:fig3_2} (c). 
\begin{figure}[t]
\centering
\includegraphics[width=0.9\columnwidth]{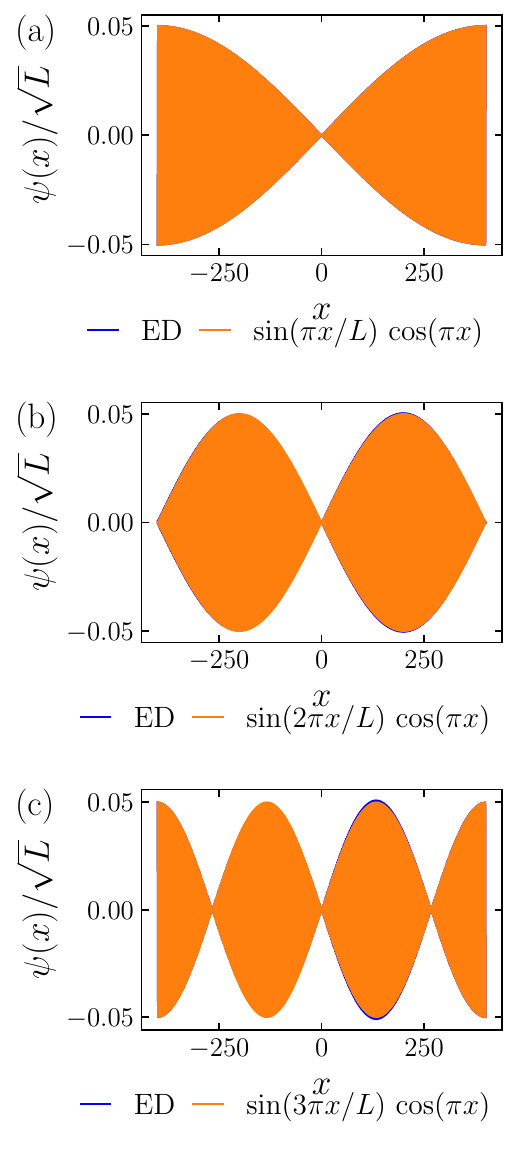}
\caption{NNN baths for $\kappa= 10, \kappa'= 5, L= 801, J=20, \Delta=2.14$. (a) shows the scattering state at the second-order pole point; (b) and (c) show the next two adjacent states. Due to the finite-size effect, they are well described by Eq.~(\ref{eq:finite-size-second-order}).}
\label{fig:fig4_2}
\end{figure}

\subsection{Second-order pole}
\label{sub:second-order}
Another case is when the $h^{\text{NNN}}_{k}$ dispersion has a vanishing derivative with respect to $k$. This is realized by a fine-tuned hopping parameter $\kappa = 2\kappa'$ and $k=\pi$. Figure \ref{fig:fig4_1} shows the spectrum, and unlike the generic parameters, the maximum winding number is one. In the thermodynamic limit, the pole is second order and given by $e^{i \pi }$. As discussed in Sec. \ref{sec:root-multi}, evaluating the finite-size results Eqs. (\ref{ex_fine}) and (\ref{ps_fine}) for the second-order pole $k_r = \pi$, the scattering wavefunction reads
\begin{equation}
    \psi_{m}(x)\simeq \cos(\pi x) \sin(\frac{m \pi x}{L}).
\label{eq:finite-size-second-order}
\end{equation}
Therefore, the closest scattering state to the second-order pole that one can get on a finite lattice is when $m=1$. And the second closest one is given by $m=2$. Fig.~\ref{fig:fig4_2} shows the exact diagonalization results, which confirm the argument above.

\section{Conclusion}
\label{sec:conclusion}
In summary, we have developed a general method for studying single-photon transport in (single-band) NH baths on finite lattices and addressed the validity of the Lippmann-Schwinger equation. For generic baths and most of the eigenvectors, our method reduces to the Lippmann-Schwinger equation in the thermodynamic limit, but it can reduce to other expressions for fine-tuned setups. The analytical solutions for the 1D Hatano-Nelson and unidirectional next-to-nearest-neighbor hopping models are computed, and we show that the scattering states are not linear superpositions of plane waves in general. 

Our work provides a solid example for studying scattering states with a complex ``kinetic term'', which is a step towards a general theory of NH scattering. In addition, our work shows some unique features of the scattering states in highly experimentally relevant optical platforms. There are several immediate directions for future work. One important theoretical question is to understand the connection with finite-size Hermitian scattering theory. Another direction is generalizing our single-band approach to multiple bands \cite{Yao2018,PhysRevResearch.7.L012036} and to higher dimensions \cite{PhysRevX.14.021011}; also, extending to the impurity problems \cite{PhysRevB.102.125124,burke2025nonhermitiannumericalrenormalizationgroup}. Another natural theoretical question is to study the strongly correlated states in the higher-excitation sectors \cite{Shen2007} for the current setup, or even in the presence of multiple emitters \cite{shi_multiphoton-scattering_2015, ShiMultimer2024}. From an application perspective, it would be interesting to investigate coherent photon transport in other experimentally accessible structured waveguides \cite{Poddubny2024}.

\begin{acknowledgments}
     Z.G. acknowledges support from the University of Tokyo Excellent Young Researcher Program and from JST ERATO Grant Number JPMJER2302, Japan. F.N. is supported in part by the Japan Science and Technology Agency (JST) [via the CREST Quantum Frontiers program Grant No. JPMJCR24I2, the Quantum Leap Flagship Program (Q-LEAP), and the Moonshot R\&D Grant No. JP-MJMS2061], and the Office of Naval Research (ONR) Global (via Grant No. N62909-23-1-2074).
\end{acknowledgments}

\typeout{}
%\bibliography{ref}

\appendix

\section{Remarks on the residue formula Eq.~(\ref{SR})}
\label{RF}
Noting that $f^{(L)}_-(y)=f^{(L)}_+(y)+1$, we know that the consistency of $\pm$ cases at $x=0$ is equivalent to
\begin{equation}
\sum_{y:E=h(y)} \frac{1}{y h'(y)}=0,
\label{sy}
\end{equation}
which simply follows from (again, $R$ is assumed to be large enough)
\begin{equation}
\oint_{|\beta|=R} \frac{d\beta}{2\pi i \beta} \frac{1}{E - h(\beta)}=0.
\label{ibr}
\end{equation}
Note that $0$ has been assumed to be not a zero of $\beta(E-h(\beta))$ or $\beta(E-h(\beta^{-1}))$, as can be easily ensured by $\min\{p,q\}\ge1$. The periodicity of $\Sigma_x^{(L)}(E)$ in $x$ can be seen from $f^{(L)}_+(y) = y^{-L} f^{(L)}_-(y)$, and that if $x\ge 0$ so that $\Sigma_x^{(L)}(E)$ involves $f^{(L)}_+(y)$, then $x-L\le 0$ and thus $\Sigma_x^{(L)}(E)$ should be evaluated using $f^{(L)}_-(y)$.

On the other hand, if $\min\{p,q\}=0$, as is the case of the NNN bath, $0$ will be a pole of $1/\left( \beta(E-h(\beta)) \right)$ or $1/\left(\beta(E-h(\beta^{-1})) \right)$ depending on the unidirectionality. In this case, applying the residue theorem to Eq.~(\ref{ibr}) (or with $h(\beta)$ replaced by $h(\beta^{-1})$) yields the following modification of Eq.~(\ref{sy}):
\begin{equation}
\frac{1}{E-h_0}+\sum_{y:E=h(y)} \frac{1}{y h'(y)}=0,
\end{equation}
provided that $E\neq h_0$. The additional term $1/(E-h_0)$ corresponds to the discrepancy between $x>0$ and $x<0$ cases of $\Sigma^{(L)}_x(E)$ at $x=0$.

\section{Discontinuity of the self-energy}
\label{app:discont-self-energy}
The self-energy $\Sigma(z)$ defined in Eq.~(\ref{eq:Green-self-energy}) has been analytically calculated for both HN and NNN models. For both examples, the self-energy shows a discontinuity in the directions of approaching $h_k$. In this section, we show that our analytical results agree with the general statement about the discontinuity in self-energies in Ref \cite{Longhi2016}. 

For a finite-hopping range, the complex dispersion $h_k$ is always continuous and differentiable. The discontinuity in self-energies is given by 
\begin{equation}
    \Sigma^{>}( h_{k }  )-\Sigma^{<}( h_{k }  ) = -\frac{i J^2}{(dh_k/dk)}.
\label{eq:discont-self-energy}
\end{equation}
For the HN case, $dh_k/dk=-i(J-\kappa/2)e^{ik}+i(J+\kappa/2)e^{-ik}=-i\delta$ and related to the discontinuity $\Sigma^{>}( h_{k})-\Sigma^{<}(h_{k} )=J^2/\delta$ via Eq.~(\ref{eq:discont-self-energy}). The same result holds for the NNN case, where $dh_k/dk=i\kappa e^{-ik}+2i\kappa'e^{-2ik}=i\delta e^{-ik}$ and $\Sigma^{>}(h^{\text{NNN}}_{ k_{1}} ) - \Sigma^{<}(h^{\text{NNN}}_{ k_{1}} ) = -J^2 e^{ik}/\delta$.

\section{Lippmann-Schwinger equation on lattices}
\label{app:lattice-LS}

In the main text, we have shown that the exact formal solution for the finite lattice reduces to the LS equation in the thermodynamic limit. Yet the LS equation itself on discrete spaces is less well known. In this section, we show how to obtain the LS equation on lattices following the idea for the continuous space \cite{Lippmann1950}.

In the zero-coupling limit ($J=0$), on top of the emitter excitation with eigenenergy $\Delta$, eigenvalues of the NH Hamiltonian in Eq.~(\ref{eq:exact-eigen}) are simply solved by plane waves 
\begin{equation}
    H_0 \left( |g\rangle \otimes  |k\rangle \right) = h_k (|g\rangle \otimes |k\rangle),
\end{equation}
where $ | k\rangle =  a^{\dagger}_k |\text{vac}\rangle$. For those scattering eigenstates, the formal solution to Eq.~(\ref{eq:exact-eigen}) at finite coupling is given by the LS equation \cite{Lippmann1950}
\begin{equation}
    |\Psi^{>,< } _k \rangle = |g\rangle \otimes  |k\rangle + (h_{k\pm i0} - H_0 )^{-1} V |\Psi^{>,<} _k \rangle, 
\label{eq:self-consistent-LS}
\end{equation}
where $<$ ($>$) indicates the branches when substituting $h_{k + i0}$ ($h_{k - i0}$), like the main text. The formal solution Eq.~(\ref{eq:self-consistent-LS}) coincides with the exact eigenvector $| \Psi_i \rangle$ in Eq.~(\ref{eq:exact-eigen}). 

In the following, we consider the Born approximation
\begin{equation}
    |\Psi^{>,<} _k \rangle \approx |g\rangle \otimes  |k\rangle + (h_{k\pm i0} - H_0  )^{-1} V |g\rangle \otimes  |k\rangle,
\label{eq:Born-LS}
\end{equation}
which truncates the self-consistent LS equation at the first-order. Noting the fact that the scattering process here conserves the particle number, we know the LS scattering wavefunction has to preserve the single excitation form
\begin{equation}
    |\Psi^{>,<} _k \rangle = |g\rangle \otimes |\psi^{>,<}_k\rangle + c_e |e\rangle \otimes |\text{vac}\rangle.
    \label{eq:se}
\end{equation}
Substituting Eq.~(\ref{eq:se}) into the Born equation in Eq.~(\ref{eq:Born-LS}) and recalling the definition of $V = \frac{J}{\sqrt{L}}\sum_k \left( |g\rangle \langle e | a^{\dagger}_k + a_k |e \rangle \langle g|   \right)$ leads to 
\begin{align}
    \begin{split}
        |\psi_k^{>,<}\rangle &= |k\rangle + \frac{J}{L} \sum_{k'}G^{>,<}_{k'}(h_{k}) |k'\rangle,\\ 
        c_e &= \frac{J}{\sqrt{L}}G^{>,<}_e(h_{k}),
    \end{split}
\end{align}
where $G_k(z)$ is the photon Green’s function
\begin{equation}
    G_k(z) = \frac{J}{z-h_k}G_e(z).
    \label{eq:Green-self-energy}
\end{equation}
Thus the wavefunction $\psi^{>,<}_{k}(x) = \sqrt{L} \langle x | \psi^{>,<}_k \rangle$ reads
\begin{equation}
    \psi^{>,<}_{k}(x) = e^{ikx} + G^{>,<}_e(h_{k})\Sigma^{>,<}_x(h_{k}).
\label{eq:LS-wavefunction}
\end{equation}

\section{Full derivations for Hatano-Nelson baths}
\label{app:derivation-HN}
In this section, we consider the full derivations of scattering wavefunction Eq.~(\ref{eq:LS-wavefunction-HN}) for HN baths in detail. We start by evaluating the self-energy Eq.~(\ref{TL}),  
\begin{align}
    \begin{split}
         \Sigma_x(z)&=\frac{J^2}{2\pi}\int^\pi_{-\pi}dk\frac{e^{ikx}}{z-h_k}\\
         &=\frac{J^2}{2\pi i}\oint_{|y|=1} \frac{dy}{y}\frac{y^{|x|}}{(z-h_k)},
    \end{split}
\label{eq:relative-self-energy-cont}
\end{align}
where the variable $k$ is changed to $y$ and $y=e^{i\text{sgn}(x)k}$. For $x<0$, $dy=-iydk$, and the substitution $y=e^{-ik}$ runs clockwise, which introduces another minus sign. The integral becomes a contour integral along the unit circle, and evaluation is done by the residue theorem. Firstly, we find the two roots $y_\pm$ of the polynomial $y(z-h_k) \equiv ay^2+by+c$. Let $\delta^2=b^2-4ac$, by the residue theorem, the relative self-energy reads
\begin{equation}
    \Sigma_x(z)=\frac{J^2}{\delta}\left(y^{|x|}_+\Theta(1-|y_+|) - y^{|x|}_-\Theta(1-|y_-|)\right),
\label{eq:relative-self-energy-HN}
\end{equation}
where $\Theta(1-|y_\pm|)$ is the Heaviside's step function.

Now we compute this integral explicitly. For $x \ge 0$, the roots of the polynomial 
\begin{equation*}
    y(z-h_k)=(u+\frac{\kappa}{2})y^2+yz+(u-\frac{\kappa}{2})=0
\end{equation*}
is given by
\begin{equation*}
    y_{\pm} = \dfrac{-z\pm\sqrt{\delta^2}}{2(u -\frac{\kappa}{2})},
\end{equation*}
where $ \delta = (u -\kappa/2)e^{ik} - (u+ \kappa/2)e^{-ik}$. So far we expressed Eq.~(\ref{eq:relative-self-energy-cont}) for an arbitrary $z$ and we consider taking the limit of $z\rightarrow h_k^{\text{HN}} \pm i0$, which reads
\begin{equation*}
y_{\pm} = \frac{(u-\frac{\kappa}{2})e^{ik}+(u+\frac{\kappa}{2})e^{-ik}\pm((u-\frac{\kappa}{2})e^{ik}-(u+\frac{\kappa}{2})e^{-ik})}{2u-\kappa}.
\end{equation*}
The roots are further simplified to 
\begin{align}
\begin{split}
     y_{+} &= e^{ik}, \\
    y_{-} &= \frac{u +\frac{\kappa}{2}}{u-\frac{\kappa}{2}}e^{-ik},
\end{split}
\label{eq:HN-roots1}
\end{align}
and the root $y_-$ always has a magnitude greater than one, unless $\kappa =0 $ which is the Hermitian limit of the HN model. Thus, by the residue theorem, $y_-$ does not contribute to Eq.~(\ref{eq:relative-self-energy-cont}) as it falls outside the unit circle. However, the root $y_+$ has a magnitude of one and is located on the integration contour, which makes the relative self-energy ill-defined. We consider an infinitesimal deformation of the integrand by considering $k \rightarrow \lim_{\epsilon\rightarrow 0 } k+i\epsilon$, where $\epsilon$ is real and positive, which changes the magnitude of the root $y_+$ to  $|e^{i(k+i\epsilon)}|=|e^{ik}e^{-\epsilon}|<1$ and resulting in pushing $e^{ik}$ into the unit circle. 

The above process is repeated for $x < 0$, and the roots of the polynomial $y(z-h_k)=0$ are given by 
\begin{equation*}
    y_{\pm}=\frac{-z\pm\sqrt{\delta^2}}{2(u +\frac{\kappa}{2})}.
\end{equation*} 
Taking the limit of $z\rightarrow h_k^{\text{HN}}$, we find 
\begin{align}
\begin{split}
    y_+ &= \frac{u -\frac{\kappa}{2}}{u+\frac{\kappa}{2}}e^{ik}, \\
    y_- &= e^{-ik}. 
\end{split}
\label{eq:HN-roots2}
\end{align}
The root $y_+$ contributes to the relative self-energy as its magnitude is less than one. Similarly, the contour integral for $y_-$ is not well-defined, and we consider the above substitution $k \rightarrow \lim_{\epsilon\rightarrow 0 } k + i\epsilon$, which changes the magnitude of $y_-$ to  $|e^{-i(k+i\epsilon)}|=|e^{-ik}e^{\epsilon}|>1$ and does not affect the integral. 
Thus, we have found the non-vanishing residues, Eq.~(\ref{eq:HN-roots1}) and Eq.~(\ref{eq:HN-roots2}), and substitute them into Eq.~(\ref{eq:relative-self-energy-HN}) give 
\begin{equation}
\Sigma^{>}_x(h^{\text{HN}}_{ k} ) =
\begin{cases}
    \frac{J^2}{\delta}e^{ikx} \quad &x\ge 0 \\
    \frac{J^2}{\delta}\left(\frac{u-\frac{\kappa}{2}}{u+\frac{\kappa}{2}}\right)^{-x}e^{-ikx} \quad &x<0.
\end{cases}
\label{eq:NN-Sigmazx-z0-}
\end{equation}

So far we have only considered the case $z = \lim_{\epsilon \rightarrow 0 } h^{\text{HN}}_k + i\epsilon$ in Eq.~(\ref{eq:HN-roots1}) and Eq.~(\ref{eq:HN-roots2}). It is remain to study the case $z = \lim_{\epsilon \rightarrow 0 } h^{\text{HN}}_k - i\epsilon$ by taking $k \rightarrow \lim_{\epsilon\rightarrow 0 } k - i\epsilon$. For $x\ge 0$, both roots Eq.~(\ref{eq:HN-roots1}) now fall outside the unit circle. And for $x < 0$, both roots Eq.~(\ref{eq:HN-roots2}) are included. The behavior of the roots leads to the following self-energy expressions 
\begin{equation}
%\Sigma_x(h^{\text{HN}}_{k + i0^{-}}) = 
\Sigma^{<}_x(h^{\text{HN}}_{ k} ) =
\begin{cases}
    0 \quad &x\ge0 \\
    \frac{J^2}{\delta}\left(\left(\frac{u-\frac{\kappa}{2}}{u+\frac{\kappa}{2}}\right)^{-x}e^{-ikx}-e^{ikx}\right) \quad &x<0.
\end{cases}
\end{equation}
The emitter Green's functions Eq.~(\ref{eq:Green-self-energy}) are given by 
\begin{align}
\begin{split}
G^{>}_e( h^{\text{HN}}_{ k} ) &= \frac{1}{h^{\text{HN}}_k -\Delta-\frac{J^2}{\delta}},\\
G^{<}_e(h^{\text{HN}}_{ k} ) &= \frac{1}{h^{\text{HN}}_k-\Delta}.
\end{split}
\label{eq:emitter-Green-HN}
\end{align}
The scattering wavefunction $\psi^{>}_{ k}(x)$ is presented in the main text. And the scattering wavefunction $\psi^{<}_{ k}(x)$ is given by 
\begin{widetext}
\begin{align}
\begin{split}
\psi^{<}_{ k}(x) =
    \begin{cases}
        e^{ikx}& x\ge0 \\
      e^{ikx} + G^{<}_e(h^{\text{HN}}_{k} ) \frac{J^2}{\delta}\left(\left(\frac{u-\frac{\kappa}{2}}{u+\frac{\kappa}{2}}\right)^{-x}e^{-ikx}- e^{ikx}\right) & x<0.
    \end{cases}
    \end{split}
\label{eq:LS-wavefunction-NH-unphysical}
\end{align}
\end{widetext}

\section{Full derivations for unidirectional next-to-nearest-neighbor baths}
\label{app:derivation-NNN}
We start by considering the relative self-energy given by Eq.~(\ref{eq:relative-self-energy-cont}). For $x \ge 0$, let $y=e^{ik}$ be the substitution and we have 
\begin{align}
    \begin{split}
          \Sigma_{x \ge 0}(z)&=\frac{J^2}{2\pi i}\oint_{|y|=1} \frac{dy}{y} \frac{y^{x+2}}{zy^2+\kappa y + \kappa'} \\
          &=\frac{J^2}{\delta}\left(y^{x+1}_+\Theta(1-|y_+|)  - y^{x+1}_-\Theta(1-|y_-|) \right),
    \end{split}
\label{eq:relative-self-energy-NNN1}
\end{align}
where $\delta=\kappa+2\kappa'e^{-ik}$ and the integral is calculated by the residue again. The roots of polynomial $zy^2+\kappa y + \kappa' = 0$ are given by 
\begin{align}
\begin{split}
    y_+ &= -\frac{\kappa'}{\kappa+\kappa'e^{-ik}}, \\
    y_- &= e^{ik}.
\end{split}
\label{eq:NNN-roots1}
\end{align}
For the case of $x<0$, the substitution becomes $y=e^{-ik}$
\begin{align}
    \begin{split}
     \Sigma_{x < 0}(z) &= \frac{J^2}{2\pi i}\oint_{|y|=1}\frac{y^{-x}}{y(\kappa'y^2+\kappa y + z)} \\
     &= \frac{J^2}{\delta}\left(y^{-x-1}_+\Theta(1-|y_+|)  - y^{-x-1}_-\Theta(1-|y_-|) \right),
\end{split}
\label{eq:relative-self-energy-NNN2}
\end{align}
we find the roots for the polynomial $\kappa'y^2+ky+z = 0$ to be
\begin{align}
    \begin{split}
    y_+ &= e^{-ik}, \\
    y_- &= -\frac{\kappa + \kappa'e^{-ik}}{\kappa'}.
\end{split}
\label{eq:NNN-roots2}
\end{align}

Recall that the $h^{\text{NNN}}_{k}$ has a self-intersecting point at 
\begin{equation*}
    k_{\text{SI}}=\arccos{(-\frac{\kappa}{2\kappa'})},
\end{equation*} 
which is a special feature of the model. The self-intersecting point makes the relative self-energy discontinuous in $k$. More precisely, the magnitude of roots in Eq.~(\ref{eq:NNN-roots1}) and Eq.~(\ref{eq:NNN-roots2}) varies as arcoss the self-intersecting point; the $|y_+|$ in Eq.~(\ref{eq:NNN-roots1}) is greater than one for  $k_1 = |k| > | k_{\text{SI}}| $ and less than one for $k_2 = |k| < | k_{\text{SI}}|$. And the opposite holds for $|y_{-}|$ in Eq.~(\ref{eq:NNN-roots2}). At the self-intersecting point, both roots $y_{\pm} = e^{\pm ik_{\text{SI}}}$ have a unit magnitude. As discussed in the main text, the exact scattering state for this point needs special treatment. 

The dispersion of NNN has two domains with non-trivial winding $1$ and $2$, unlike the case of HN model where the maximum winding number is $1$. Therefore, one has to specify the domain of $z$ when taking the following limit $ z\rightarrow h_k^{\text{NNN}} \pm i0$ in the explicit evaluation of all the Green's functions and self-energies Eq. (\ref{TL}). The winding number of $h_k^{\text{NNN}}$ is $1$ or $2$ for $k_{1/2}$, and it is useful to label the dispersion $h_{k_{1/2}}^{\text{NNN}}$ as well.

Now we consider the contributions of the residues to the relative self-energy in Eq.~(\ref{eq:relative-self-energy-NNN1}) and  Eq.~(\ref{eq:relative-self-energy-NNN2}). In the winding number $1$ domain, where the momentum is strictly defined by $k=k_1$. We first consider the case $k \rightarrow \lim_{\epsilon\rightarrow 0 } k+i\epsilon$, where $\epsilon$ is real and positive. Again, this affects the magnitude of the root $e^{ik}$ and change it to $|e^{ik}e^{-\epsilon}|<1$. The self-energies are  
\begin{equation}
    \Sigma^{>}_x(h^{\text{NNN}}_{ k_{1}} ) = 
    \begin{cases}
        \frac{J^2}{\delta}\left((-\frac{\kappa'}{\kappa+\kappa'e^{-ik_1}})^{x+1}-e^{ik_1(x+1)}\right) & x \ge 0 \\
        % J^2\frac{e^{ik}}{\kappa + \kappa'e^{-ik}} & x=0 \\
        0 & x<0.
    \end{cases}
\end{equation}

Similarly, we also consider the solutions by taking the limit $k \rightarrow \lim_{\epsilon\rightarrow 0 } k- i\epsilon$ from the negative side. We find the self-energies to be
\begin{equation}
    \Sigma^{<}_x(h^{\text{NNN}}_{ k_{1}} ) = 
    \begin{cases}
        \frac{J^2}{\delta}(-\frac{\kappa'}{\kappa+\kappa'e^{-ik_1}})^{x+1} & x \ge 0 \\
        \frac{J^2}{\delta}e^{ik_1(x+1)} & x<0.
    \end{cases}
\end{equation}
We repeat the above evaluations for the relative self-energy in the winding number $2$ domain, where the momentum is $k=k_2$. The following two sets of solutions correspond to the limits $k \rightarrow \lim_{\epsilon\rightarrow 0 } k \pm i\epsilon$. In the case of $+$, the self-energies are
\begin{equation}
    \Sigma^{>}_x(h^{\text{NNN}}_{ k_{2}} ) = 
    \begin{cases}
        -\frac{J^2}{\delta}e^{ik_2(x+1)} & x \ge 0 \\
        -\frac{J^2}{\delta}(-\frac{\kappa'}{\kappa+\kappa'e^{-ik_2}})^{x+1} & x<0.
    \end{cases}
\end{equation}
Lastly, we have the self-energies for $-$
\begin{equation}
    \Sigma^{<}_x(h^{\text{NNN}}_{ k_{1}} ) = 
    \begin{cases}
        0 & x \ge 0 \\
        \frac{J^2}{\delta}\left(e^{ik_2(x+1)}-(-\frac{\kappa'}{\kappa+\kappa'e^{-ik_2}})^{x+1}\right) & x<0.
    \end{cases}
\end{equation}
In the NNN case, the emitter Green's functions Eq.~(\ref{eq:Green-self-energy}) are discontinuous at $x=0$, and they are 
\begin{align}
\begin{split}
G^{>,x\ge 0}_e( h^{\text{NNN}}_{ k_{1}}  ) &=
\frac{1}{h^{\text{NNN}}_{k_1} -\Delta + J^2(\frac{e^{ik_1}}{\kappa+\kappa'e^{-ik_1}})},  \\
G^{>,x<0}_e( h^{\text{NNN}}_{ k_{1}}  ) &= \frac{1}{h^{\text{NNN}}_{k_1} -\Delta }, \\
G^{<,x\ge0}_e( h^{\text{NNN}}_{ k_{1}}  ) &= \frac{1}{h^{\text{NNN}}_{k_1} -\Delta  +\frac{J^2}{\delta}(\frac{\kappa'}{\kappa+\kappa'e^{-ik_1}})},  \\
G^{<,x<0}_e( h^{\text{NNN}}_{ k_{1}}  ) &= \frac{1}{h^{\text{NNN}}_{k_1} -\Delta  - \frac{J^2}{\delta}e^{ik_1}},  \\
G^{>, x\ge0}_e( h^{\text{NNN}}_{ k_{2}}  ) &= \frac{1}{h^{\text{NNN}}_{k_2} -\Delta  +\frac{J^2}{\delta}e^{ik_2}},  \\
G^{>,x < 0}_e( h^{\text{NNN}}_{ k_{2}}  ) &= \frac{1}{h^{\text{NNN}}_{k_2} -\Delta -\frac{J^2}{\delta}(\frac{\kappa'}{\kappa+\kappa'e^{-ik_2}})},  \\
G^{<,x\ge0}_e( h^{\text{NNN}}_{ k_{2}}  ) &= \frac{1}{h^{\text{NNN}}_{k_2} -\Delta },  \\
G^{<,x<0}_e( h^{\text{NNN}}_{ k_{2}}  ) &= \frac{1}{h^{\text{NNN}}_{k_2} -\Delta -J^2(\frac{e^{ik_2}}{\kappa + \kappa'e^{-ik_2}})}.
\end{split}
\label{eq:emitter-Green-NNN}
\end{align}
In additional to the two sets of scattering wavefucntions $\psi^{>}_{k_{1/2}(x)}$ present in the main text, the explicit form of $\psi^{<}_{k_{1/2}}(x)$ read as 
\begin{widetext}
\begin{align}
\begin{split}
&\psi^{<}_{k_1 }(x) = 
    \begin{cases}
        e^{ik_1 x}+G^{x\ge 0}_e( h^{<,\text{NNN}}_{ k_{1}}  )\frac{J^2}{\delta}(-\frac{\kappa'}{\kappa+\kappa'e^{-ik_1}})^{x+1} & x\ge 0 \\
        e^{ik_1 x}+G^{x<0}_e( h^{<,\text{NNN}}_{ k_{1}} )\frac{J^2}{\delta}e^{ik_1(x+1)} & x<0,
    \end{cases}    \\
&\psi^{<}_{k_2}(x) = 
    \begin{cases}
       e^{ik_2 x} & x\ge 0 \\
       e^{ik_2 x}+G^{x<0}_e( h^{<,\text{NNN}}_{ k_{2}}  )\frac{J^2}{\delta}\left(e^{ik_2 (x+1)}-(-\frac{\kappa'}{\kappa+\kappa'e^{-ik_2}})^{x+1}\right) & x<0.
    \end{cases}     
    \end{split}
\label{eq:LS-wavefunction-NNNother}
\end{align}
\end{widetext}

\bibliography{scatter-ref-capitalized}

\end{document}